\documentclass[aps,showpacs,amsmath,amssymb]{revtex4}
\usepackage{graphicx}

\usepackage{graphicx}
\usepackage{bm}

\begin{document}


\title{The thermodynamic cost of driving quantum systems by their boundaries}
\author{Felipe Barra\footnote{fbarra@dfi.uchile.cl}}
\affiliation{Departamento de F\'isica, Facultad de Ciencias F\'isicas y Matem\'aticas, Universidad de Chile, Santiago Chile}

\begin{abstract}
The laws of thermodynamics put limits to the efficiencies of thermal machines. Analogues of these laws are now established for quantum engines weakly and passively coupled to the environment providing a framework to find improvements to their performance. 
Systems whose interaction with the environment is actively controlled do not fall in that framework. 
Here we consider systems actively and locally coupled to the environment, evolving with a so-called boundary-driven Lindblad equation. 
Starting from a unitary description of the system plus the environment we simultaneously obtain the Lindblad equation and the appropriate expressions for heat, work and entropy-production of the system 
extending the framework for the analysis of new, and some already proposed, quantum heat engines. We illustrate our findings in spin 1/2 chains and explain why an XX chain coupled in this way to a single heat bath relaxes to thermodynamic-equilibrium while and XY chain does not. Additionally, we show that an XX chain coupled to a left and a right heat baths behaves as a quantum engine, a heater or refrigerator depending on the parameters, with efficiencies bounded by Carnot efficiencies. 
\end{abstract}

\maketitle
%
%

\section*{Introduction}

Considerable experimental progress in various physical systems has been achieved toward the goal of controlling the dynamics of open quantum systems and their interactions with the environment~\cite{exp-gen,exp-gen1,exp-gen2}.  
For quantum computations or digital coherent quantum simulations, one may wish to have a system that is well isolated from the environment. 
For dissipative variants of quantum computations~\cite{Cirac} or creating new scenarios for non-equilibrium many-body systems, one would need to engineer the coupling to the environment.
Recently, a
setting in which the quantum system of interest interacts at its boundaries with an external quantum probe
such that their coupling can be localized and can be switched on and
off repeatedly with a controlled and well-defined state for the probe prior to the interaction has been experimentally realized~\cite{ion-trap}.  
This repeated interaction scheme has also been theoretically studied~\cite{AP,Karevski-Platini}. Importantly, the dynamics in an appropriate limit is a boundary-driven Lindblad equation.
In this article, we explore the question of what is the thermodynamic cost of having such operations on an open quantum system and what are the thermodynamical quantities, such as heat and work that will determine the efficiency of quantum engines operating in this manner. 
Boundary-driven Lindblad equations have been intensively studied theoretically, particularly for one-dimensional quantum chains~\cite{Karevski-Platini,Wichterich07, prosenXY,prosenXY2,prosenXY3,marko,marko2, integrable,integrable2,integrable3,integrable4,integrable5,symm}, and powerful techniques have been developed to find their non-equilibrium steady states (NESS)~\cite{marko,marko2, integrable,integrable2,integrable3,integrable4,integrable5,symm}. 
These equations are also frequently used to describe quantum engines~\cite{mari,linden,restrepo,bren} and other complex open
quantum systems coupled to one or several environments~\cite{Bouko,AB2013,Shigeru,norditaUs} because they are easy to implement.  
Nevertheless, a boundary driven Lindblad equation does not correctly describe a quantum system passively and weakly coupled to
a heat-bath as often occurs in natural systems. It was pointed out recently~\cite{kosloff} that inconsistencies with the second law of thermodynamics may arise in this case
and a careful examination of the coupling between a quantum refrigerator and the heat-baths~\cite{alonso1} reveals why boundary driven models are inappropriate for these situations.  
For a system passively and weakly coupled to one or several heat-baths
the master equation derived in the Born-Markov-secular approximation~\cite{BreuerBook} yields a proper description of the system and the correct balance of heat flows and irreversible entropy production.

Thus, for our study, we consider explicitly the active (time-dependent) type of interaction between the system and the environment implemented in~\cite{ion-trap} and the model developed in~\cite{AP,Karevski-Platini}. We apply the results of ~\cite{esposito-NJP,Reeb-Wolf} to derive the appropriate thermodynamical quantities and, in particular, we focus in the limit where the system is described by a boundary driven Lindblad equation.

Our main result is that 
driving at the boundaries, even though it looks like a work-free operation, actually might bring work to the system.
We illustrate our findings on boundary-driven spin 1/2 chains coupled to one or two heat baths. 
We show that an XX spin 1/2 chain coupled in this way to a single heat bath relaxes to
thermodynamic equilibrium while an XY does not because it is driven out of equilibrium by the power produced by the coupling to the heat bath. 
When two baths are connected to the chain, we observe that for different parameters, 
the chain operates as a quantum heat engine, refrigerator or heater, and we determine their efficiencies in the simple case of a chain of two sites. 
The rest of the paper is organized as follows. We start by reviewing first the thermodynamics of Markovian open quantum systems in the weak coupling limit~\cite{alicki,Spohn,Spohn-Leb,qtermo} and second 
a formulation~\cite{esposito-NJP} where the ``universe", system plus the environment, evolves unitarily. 
After that, we consider the repeated interaction scenario for the system and the environment from which the boundary-driven Lindblad equation and the appropriate thermodynamical quantities for the open system are obtained. 
Then we illustrate our results in XX and XY spin 1/2 chains and offer our conclusions. Finally we have collected in section Methods some details of the calculations.

\section*{Thermodynamics of open quantum systems}
\label{sec2}

\subsection*{open system weakly and passively coupled to the environment:}
Let us briefly review the usual formulation of thermodynamics in open quantum systems \cite{alicki,Spohn,Spohn-Leb,qtermo}. 
Consider an open system described by a master equation in the Lindblad form
\begin{equation}
\frac{d\rho_S}{dt}=-i[H_S(t), \rho_S]+\sum_r{\mathcal D}_r( \rho_S)
\label{Lind}
\end{equation}
where the environment consists of several heat-baths $r$ whose action on the system is represented by the dissipator
\[
{\mathcal D}_r(\rho)=\sum_\mu\gamma^\mu_r( 2L^\mu_r\rho L^{\mu\dag}_r-\{L^{\mu\dag}_r L^\mu_r,\rho\})
\]
with $[\cdot,\cdot]$ the commutator and $\{\cdot,\cdot\}$ the anti-commutator.
The operators $L^\mu_r$ are system operators and represent the action of the environment over the system. 
When this equation is obtained from the weak coupling limit for a time independent system, one finds {\it global} Lindblad operators $L^\mu_r$ that are eigen-operator of the Hamiltonian $H_S$~\cite{BreuerBook}.
For simplicity, we consider that the system can only exchange energy and no particles with the environment. 

Now consider the internal energy $U(t)={\rm tr}(H_S(t)\rho_S(t))$ and the entropy $S(t)=-k_B{\rm Tr}(\rho_S(t)\ln\rho_S(t))$. 
The first law of thermodynamics $\dot U=\dot W+\dot Q$ splits the rate of change of internal energy in two, power $\dot W(t)$ and heat flow $\dot Q(t)=\sum_r\dot Q_r(t)$
with one contribution per heat-bath.
For system passively and weakly coupled to the heat-baths, these quantities are defined as 
\begin{equation}
\dot W(t)={\rm tr}[\dot H_S(t)\rho_S(t)],\quad\dot Q_r(t)={\rm tr}[H_S(t){\mathcal D}_r(\rho_S(t))].
\label{WQwcl}
\end{equation} 
In section {\it Methods: Heat from a given reservoir in the weak-coupling limit} we justify these definitions.
Note that if the Hamiltonian of the system is time independent, no work can be performed on the system and only heat is exchanged with the baths. In that case the system will typically reach a steady state. 
Consider now this to be the situation. 
The second law states the positivity of the entropy production ($d_iS/dt\geq 0$), which is the difference between the time-derivative of the entropy $\dot S(t)=-\sum_r k_B{\rm Tr}({\mathcal D}_r( \rho_S)\ln\rho_S(t))$ and the entropy flow from the environment to the system $\sum_r\beta_r \dot Q_r$,  
\begin{equation}
\frac{d_i S}{dt}=\dot S-\sum_r\beta_r \dot Q_r=-k_B\sum_r{\rm Tr}({\mathcal D}_r(\rho_S)(\ln\rho_S-\ln\omega_{\beta_r}(H_S))).
\label{eprod0}
\end{equation}
The canonical distribution $\omega_{\beta_r}(H)=\exp(-\beta_r H)/Z_r$ appears in the last equality of Eq.(\ref{eprod0}) due to the definition of heat that we plug in the first equality in Eq.(\ref{eprod0}). 
The second law $d_i S/dt\geq 0$ in Eq.(\ref{eprod0}) holds 
if for every $r$, $\dot\rho_S=-i[H_S, \rho_S]+{\mathcal D}_r( \rho_S)$ relaxes towards the unique equilibrium state $\omega_{\beta_r}(H_S)$. 
This is the local-detailed-balance condition~\cite{espositoPRE} i.e. if a single heat-bath is in contact with the system detailed balance as defined in~\cite{Spohn,AlickiDB} holds.
This property of the dissipators ${\mathcal D}_r$ is satisfied
in quantum master equations obtained in the weak-coupling and with the Born-Markov-secular approximation (global Lindblad equation). This framework has been applied successfully 
to the study of thermodynamic properties and efficiencies of engines~\cite{alonso1,alonso2,kosloff-ok1,kosloff-ok2}.

In boundary-driven systems the Lindblad operators $L^\mu_r$ act locally on the boundaries of the system and in general the corresponding Lindblad equation does not satisfy local-detailed-balance. We come back to this point later.
Following recent developments in the physics of non-equilibrium systems that have emphasized the importance of time reversal symmetry at the microscopic level of description~\cite{campisi},
a formulation of quantum thermodynamics in which the system plus the environment evolves unitarily has been proposed~\cite{esposito-NJP}. We consider this framework to analyze boundary driven systems.


\subsection*{``universe" under unitary evolution}
\label{sec.esposito}

Let a system and an environment with Hamiltonians $H_S(t)$ and
$H_E$ (time independent), respectively, 
coupled by an interaction potential $V(t)$
evolve with total Hamiltonian $H_{\rm tot}(t)=H_S(t)+H_E+V(t)$. The environment might
consists of several heat baths $\rho_E=\bigotimes_r \omega_{\beta_r}(H_r)$
with $\omega_{\beta_r}(H)=e^{-\beta H_r}/Z_r$ the initial density matrix for the reservoir $r$. 
Initially, the system and heat baths are uncorrelated
$\rho_{\rm tot}(0)=\rho_S(0)\otimes \rho_E$. 
For arbitrary strength coupling between the system and environment~\cite{esposito-NJP},
the internal energy is defined by
$E(t)\equiv{\rm Tr}(\rho_{\rm tot}(t)(H_S(t)+V(t))),$ 
and the first law relates its changes to work and heat
$\Delta E(t)=W(t)+Q(t)$
with the work $W(t)\equiv{\rm
  Tr}(\rho_{\rm tot}(t)H_{\rm tot}(t)-\rho_{\rm tot}(0)H_{\rm tot}(0))$ performed on
the system in the time interval $[0,t]$, which is also given by
\begin{equation}
W(t)=\int_0^t dt' {\rm Tr}[\rho_{\rm tot}(t')(\dot H_S(t')+\dot V(t')))]
\label{work}
\end{equation}
and the total heat flow 
$Q(t)=\sum_rQ_r(t)$
split in reservoir contributions
\begin{equation}
Q_r(t)={\rm Tr}(H_r\rho_{\rm tot}(0))-{\rm Tr}( H_r\rho_{\rm tot}(t))
\label{Qr}
\end{equation}
given by minus the change in energy of the $r$-reservoir. 

Considering 
$S(t)=-k_B{\rm Tr}_S(\rho_S(t) \ln \rho_S(t))$ as the thermodynamic entropy of the system
and $\Delta S(t) \equiv S(t)-S(0)$, it is found that
$\Delta S(t) =\Delta_{\rm e} S(t) + \Delta_{\rm i} S(t)$
with the entropy flow $\Delta_{\rm e} S(t)=\sum_r\beta_r Q_r(t)$ determined by the heat flows in Eq.(\ref{Qr})
and the entropy production~\cite{esposito-NJP}
\begin{equation}
\Delta_{\rm i} S(t)=D(\rho_{\rm tot}(t)||\rho_S(t)\otimes \rho_E)
\geq 0,
\label{ep}
\end{equation}
with $D(a||b)={\rm Tr}(a\ln a)-{\rm Tr}(a\ln b)$. Unitarity, expressed through the invariance of ${\rm Tr}(\rho_{\rm tot}(t)\ln\rho_{\rm tot}(t))$ under the time evolution of the full system, plays a crucial role in the splitting of entropy change in the entropy flow and a positive entropy production. 
In the weak-coupling limit $V\to 0$
and assuming that the open system satisfies a Lindblad equation obtained from the Born-Markov-secular approximation~\cite{BreuerBook},
the rate of entropy production $d_{\rm i}S/dt\geq 0$ and the above expressions for work and heat take the standard form given in Eq.(\ref{eprod0}) and Eq.(\ref{WQwcl}) respectively.  
This is shown in section methods by considering the method of full-counting statistics~\cite{esposito-RMP}.
However, the Lindblad models 
investigated in~\cite{kosloff, Karevski-Platini,prosenXY,prosenXY2,prosenXY3, integrable,integrable2,integrable3,integrable4,integrable5,marko,marko2} are not obtained from the weak-coupling limit and do not satisfy local-detailed-balance.
Thus to obtain the appropriate expressions for the thermodynamical quantities in boundary driven systems we apply in the next section the previous formulation, in particular Eqs.(\ref{work},\ref{Qr},\ref{ep}), to a system plus environment evolving unitarily in which the reduced density matrix for the system satisfy a boundary driven Lindblad equation in an exact limit.

\section*{The repeated interaction scheme} 
\label{sec3}
Let us consider a finite system with time-independent Hamiltonian $H_S$ and left $(L)$ and right $(R)$ reservoirs composed of an infinite set of identical non-interacting finite systems with Hamiltonian $H_r^n$, i.e., $H_{r}=\sum_nH_r^n$, where $r$ is $L$ or $R$. 
Each $H_r^n$ interacts with the system for a time span $\tau$. This interaction is always of the same form, but to emphasize that interactions occur with different copies $H_L^n+H_R^n$ in different time intervals, we write it as $V(t)=V^{n}$  if $t\in [(n-1)\tau,n\tau]$ with $V^{n}=V_L^{n}+V_R^{n}$. At $t=0$, the system and reservoirs are decoupled, i.e., $\rho_{\rm tot}=\rho_S(0)\otimes \rho_E$, with $\rho_S(0)$ arbitrary and $\rho_E=\bigotimes_{n} \rho_n$, where $\rho_n=\omega_{\beta_L}(H_L^n) \otimes \omega_{\beta_R}(H_R^n)$. At $t=0^+$, the system begins to interact with the first copy $H_L^1+ H_R^1$, and after a lapse of time $\tau$, the state of the total system is $\rho_{\rm tot}(\tau)=U_1[\rho_S(0)\otimes\rho_1]U_1^\dag\otimes \rho_2\otimes\rho_3...$. Then, at $t=\tau+0$, 
the interaction with the first copy is replaced by an interaction with
the second copy for a time $\tau$ and so on. A recursion relation for
the state of the system is obtained~\cite{AP,Karevski-Platini} by
tracing out the $n$th copy of the environment (denoted as ${\rm Tr}_n$)
\begin{equation}
\rho_S(n\tau)={\rm Tr}_{n}[U_n(\rho_S((n-1)\tau)\otimes \rho_n)U_n^\dag].
\label{rint}
\end{equation}
The unitaries are $U_n=e^{-i\frac{\tau}{\hbar}(H_S+H^{n}_L+H^{n}_R+V^{n})}$. This is the repeated interaction scheme. For simplicity we considered only two heat-baths but the generalization to several reservoirs is straight forward.

Let us consider
the change of thermodynamical quantities 
in the time intervals of length $\tau$. 
Crucially, due to the resetting of the heat baths, the interaction term is time dependent. 
According to Eq.(\ref{work}) for time-independent $H_S$, work is performed at the discrete times $n\tau$ where the interaction between the system and the environment changes because the copy in interaction changes. 
Performing the integral in Eq.(\ref{work}) between an initial time $n\tau-\epsilon$ and a final time $n\tau+\epsilon$, we obtain 
$\Delta W_{n\tau}={\rm Tr}_{\rm tot}([V^{n+1}-V^n]\rho_{\rm tot})$ in the limit $\epsilon\to 0$.
We simplify this expression with the standard~\cite{BreuerBook} assumption that ${\rm Tr}_r(V_r^{n} \omega_{\beta}(H_r^n))=0$. 
This condition will be repeatedly used; it allows us to split $\Delta W_{n\tau}=\Delta W_L+\Delta W_R$ (we drop the index $n\tau$) with
\begin{equation}
\Delta W_r=-{\rm Tr}(V^n_rU_n\rho_S([n-1]\tau)\otimes \rho_{n}U_n^\dag).
\label{WRI}
\end{equation}
We use ${\rm Tr}_r$ to denote the trace over the $r=L$ or $r=R$ system
and ${\rm Tr}$ to denote the full trace. 

The heat flow from the bath to the system in the time interval of length $\tau$ where the system interacts with the $n$th copy is evaluated from Eq.(\ref{Qr}) 
\begin{equation}
\Delta Q_r
={\rm Tr}(H_r^n [\rho_n-\rho_{n}']),
\label{QRI}
\end{equation}
where $\rho'_{n}={\rm Tr}_S(U_n\rho_{S}([n-1]\tau)\otimes \rho_n U_n^\dag)$ is the density matrix of the $n$th copy of the environment at the end of the interaction with
the system. 

The entropy production $\Delta_{\rm i}S$ in the time lapse $\tau$ is obtained from Eq.(\ref{ep}),  
and after some manipulations~\cite{esposito-NJP,Reeb-Wolf}, it can be written as the sum
\[
\Delta_{\rm i}S=D(\rho_n'||\rho_n)+I(S':n')\geq 0
\]
where the mutual information $I(S':n')\equiv S(\rho_S(n\tau)+S(\rho'_{n})-S(U_n\rho_{S}([n-1]\tau)\otimes \rho_n U_n^\dag)$ 
quantifies the correlations built up between the system and the $n$th copy after time $\tau$.
Note that $D(\rho_n'||\rho_n)\geq 0$ and $I(S':n')\geq 0$ and vanishing entropy production requires $\rho_n'=\rho_n$
and the absence of correlations between the system and the copy $I(S':n')=0$.  
Note that because before the interaction the state of the system is arbitrary and uncorrelated with the
product of thermal states of the copy, the theory of~\cite{esposito-NJP,Reeb-Wolf} applies independently of the correlations built 
between the system and previous copies.

\subsection*{Heat, work and boundary-driven Lindblad equation} 
The index $n$ is associated with the copy that
interacts in the interval of time $[(n-1)\tau,n\tau]$, but the copies
are all identical prior to the interaction
(a tensor product of two canonical distributions)
and the interaction $V^n_r$ is always of the same form. 
Because no confusion will arise, we drop the label $n$ and denote the
interaction $V=\sum_r V_r$, the Hamiltonian of the bath copy $H_r$ and the state
$\rho_n=\omega_{\beta_L}\otimes\omega_{\beta_R}$ with $\omega_{\beta_r}\equiv\omega_{\beta_r}(H_r)$.
It was shown~\cite{AP,Karevski-Platini} that for $V_r$ that satisfies ${\rm Tr}_r[V_r\omega_{\beta_r}]=0$ and whose strength is scaled with $\tau$ as $V_r=v_r/\sqrt{\tau}$, the system evolution Eq.(\ref{rint}) in the limit $\tau\to 0$ 
converges to a Lindblad evolution (see methods)
\begin{equation}
\dot \rho_S=-i[H_S,\rho_S]+\sum_r{\mathcal D}_r(\rho_S)
\label{BD2R}
\end{equation}
with ${\mathcal D}_r(\rho_S)={\rm Tr}_{r}[v_r(\rho_S\otimes
  \omega_{\beta_r})v_r]-\frac{1}{2}{\rm Tr}_r\{v_r^2,\rho_S\otimes
\omega_{\beta_r}\}$. 
This equation applied to particular systems provides boundary-driven Lindblad equations.

Consider now $\dot W_r=\Delta W_{r}/\tau$ and $\dot Q_r=\Delta Q_r/\tau$ with $\Delta W_r$ in Eq.(\ref{WRI}) and $\Delta Q_r$ in Eq.(\ref{QRI}). In the limit $\tau\to 0$ with $V=v/\sqrt{\tau}$,
we obtain (see methods)
\begin{equation}
\dot W_r=D_r(H_S+H_r),\quad \dot Q_r=-D_r(H_r)
\label{dotQ}
\end{equation}
where $D_r(A)={\rm Tr}[(v_rAv_r-\frac{1}{2}\{v_r^2,A\})\rho_S(t)\otimes\omega_{\beta_r}]$. 
Note the first law $\sum_r(\dot Q_r +\dot W_r)= \langle \dot H_S\rangle_t$, where 
$ \langle\dot H_S\rangle_t={\rm Tr}_S(H_S\dot \rho_S(t))=\sum_rD_r(H_S).$  
Finally, we express the entropy production rate as the difference between the time derivative of the von Neumann entropy and the entropy flow 
\begin{equation}
\frac{d_{\rm i} S}{dt}=-{\rm Tr}_S({\mathcal D}(\rho_S(t))\ln \rho_S(t))-\sum_r \beta_r \dot Q_r\geq 0
\label{dotSi}
\end{equation}
where the first term is computed using Eq.(\ref{BD2R}) with ${\mathcal D}\equiv\sum_r {\mathcal D}_r$ and the second
term is computed from Eq.(\ref{dotQ}).
Eqs.(\ref{dotQ},\ref{dotSi}) provide appropriate thermodynamic expressions for systems evolving with Eq.(\ref{BD2R}). Now we illustrate our findings in spin 1/2 chains.

\section*{Spin models} 
\label{secSpin}
Consider an XY spin 1/2 chain with Hamiltonian  
\begin{equation}
H_S=\frac{1}{2}\sum_{j=1}^N h_j\sigma_j^z-\sum_{j=1}^{N-1}(J_x\sigma^x_j\sigma^x_{j+1}+J_y\sigma^y_j\sigma^y_{j+1}).
\label{XYH}
\end{equation}
In the repeated interaction scheme we consider the couplings
\begin{equation}
V_L=J_{L}( \sigma_{L}^x \sigma_1^x+ \sigma_{L}^y \sigma_1^y),\, V_R= J_{R}( \sigma_{R}^x \sigma_N^x+ \sigma_{R}^y \sigma_N^y)
\label{XXCoup}
\end{equation}
to a left $r=L$ and a right $r=R$ spin 1/2 reservoir copy with Hamiltonians $H_r=h_r/2\sigma_r^z$,
and we take $h_L=h_1$ and $h_{R}=h_N$.
To obtain the boundary-driven Lindblad model, we scale $J_r=\sqrt{\lambda_r/\tau}$.
The canonical density matrices $\omega_{\beta_r}$
are fully characterized by the magnetization $M_r={\rm Tr}(\sigma_r^z\omega_{\beta_r})=-\tanh(\beta_r h_r/2)$.

Evaluating the second term on the right-hand side of Eq.(\ref{BD2R}) yields the dissipator in the Lindblad from
${\mathcal D}_r(\rho)=\sum_{\mu\in\{+,-\}}\gamma^{\mu}_r[2L^\mu_r\rho L^{\mu\dag}_r-\{L^{\mu\dag}_r L^\mu_r,\rho\}]$ 
with $\gamma^\pm_r=\lambda_r(1\pm M_r), L^{\pm}_L=\sigma_1^{\pm}$ and $L_R^{\pm}=\sigma_N^{\pm}$ where $\sigma^\pm_j\equiv(\sigma^x_j\pm i\sigma^y_j)/2$.
Note that $\gamma^+_r/\gamma^-_r=e^{-\beta_r h_r}$. 

This system does not satisfy local-detailed-balance with respect to the Gibbs state, i.e. $\omega_{\beta_r}=e^{-\beta_r H_S}/Z_r$ is not the solution of $0=-i[H_S,\rho]+{\mathcal D}_r(\rho)$ with $r$ either $R$ or $L$ because  ${\mathcal D}_r(\omega_{\beta_r})\neq 0$.
What can be shown is that these dissipators thermalize the single spin in the boundary if we disconnect it from the rest of the chain. Indeed let us consider the $L$ dissipator
\[
{\mathcal D}_L(\rho)=\gamma^{+}_L\left([2\sigma_1^+\rho \sigma_1^{-}-\{\sigma_1^{-} \sigma_1^+,\rho\}]+e^{\beta_Lh_1}[2\sigma_1^-\rho \sigma_1^{+}-\{\sigma_1^{+} \sigma_1^-,\rho\}]\right)
\]
upon evaluation we see that ${\mathcal D}_L(e^{-\beta_L h_1\sigma^z_1/2})=0$. This is the generic situation in boundary driven Lindblad systems. 

The expression for power and heat Eq.(\ref{dotQ}) can be evaluated using the system hamiltonian Eq.(\ref{XYH}), the coupling Eq.(\ref{XXCoup}), the
bath hamiltonian $h_r\sigma^z_r/2$ and the corresponding $\omega_{\beta_r}$. One obtain (we take $\lambda_L=\lambda_R=\lambda$)
\begin{equation}
\dot Q_L= 2h_L \lambda (M_L-{\rm Tr}_S(\sigma_1^z\rho_S(t)))
\label{QXY}
\end{equation}
and 
\begin{equation}
\dot W_L=2 \lambda  {\rm Tr}_S((J_x\sigma_1^x\sigma_{2}^x+J_y\sigma_1^y\sigma_{2}^y)\rho_S(t)).
\label{WXY}
\end{equation}
Replacing the indices $\{L,1,2\}$ by $\{R,N,N-1\}$ in Eqs.(\ref{QXY},\ref{WXY}) one has the corresponding $\dot Q_R$ and $\dot W_R$.
To compute this quantities, we obtain $\rho_S(t)$ by solving the Lindblad equation~\cite{prosen08}. 

Consider the case in which the system interacts with one bath (for instance the left bath, but we drop the label $L$).
In general, two situations can occur: the system relaxes to thermodynamic equilibrium 
in which all current vanishes or the system reaches a NESS if it is externally driven.

\begin{figure}[h]
\begin{center}
\includegraphics[
height=2.0in, width=3.0in ] {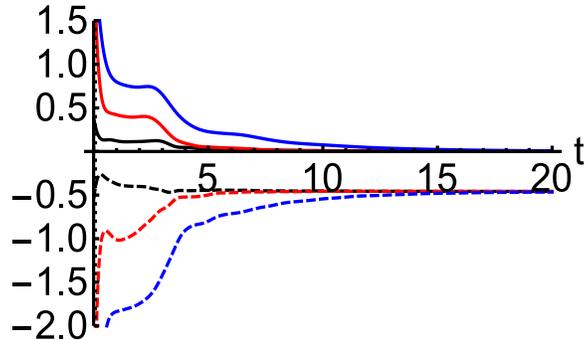}
\caption{As a function of time $t$ plots of $\dot W$ (blue), $-\dot Q$ (black) and $d_{\rm i}S/dt$ (red) for an XX ($J_x=1=J_y$) and $-\dot W$ (blue, dashed), $+\dot Q$ (back, dashed) and $-d_{\rm i}S/dt$ (red, dashed) for an XY ($J_x=1=0.5 J_y$) chain. In both cases, the chain has $N=5$ sites with $h_i=1$, $i=1,5$ coupled with $\lambda=1$ to a single left bath of $\beta=1$ and $h=1$.}
\label{Onebath}
\end{center}
\end{figure}

{\it XX chain coupled to one bath}: An XX spin chain ($J_x=J_y$) in a uniform magnetic field $h_i=h$
coupled to a single bath relaxes to equilibrium: the entropy production rate, heat flows and power vanish. 
The equilibrium density matrix is not generally a canonical
distribution but rather, as one can prove, 
is given by a generalized Gibbs state $\omega_{\beta}(H_0)$
with $H_0=\frac{h}{2}\sum_{j=1}^N \sigma_j^z$, which is a conserved quantity, i.e., $[H_S,H_0]=0$. 
This state is a product state of the canonical density matrices $\omega_\beta$ for each spin of the chain 
and all equal to the one of the reservoir copy. Therefore, $I(S':n')=0$ and $\rho_n'=\rho_n$, i.e., $d_{\rm i}S/dt=0$. 
Figure~\ref{Onebath} illustrates the relaxation to this equilibrium state 
by depicting the decaying power, heat flow and entropy production rate.

{\it XY chain coupled to a single bath}:
For an XY chain, we found that the system reaches
a driven NESS. In this NESS, 
entropy production is strictly positive and constant, 
and because $\langle\dot H_S\rangle=0$, the first law gives $\dot
W=-\dot Q$. Furthermore, by combining the first and second laws, we have that $\beta\dot W=d_{\rm i} S/dt>0$ because in NESS, $\dot S=0$. 
See Figure \ref{Onebath}. 

\begin{figure}[h]
\begin{center}
\includegraphics[
height=2.0in, width=3.0in ] {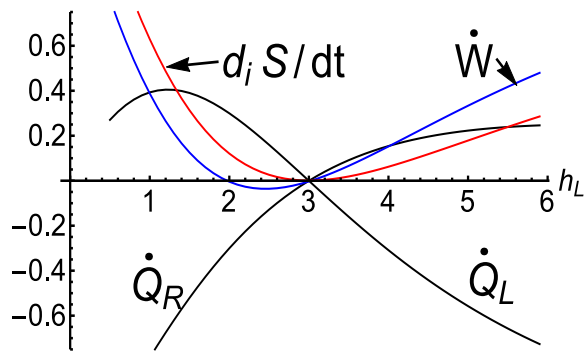}
\caption{For a $N=5$ site XX chain with $J_x=J_y=3$, $h_2=h_3=h_4=5$, $h_5=h_R=2$, $\beta_L=0.8,\beta_R=1.2$, 
and $\lambda=1$, we depict  
$\dot Q_L^{\rm NESS},\dot Q_R^{\rm NESS}$, $\dot W^{\rm NESS}$ and $(d_iS/dt)^{\rm NESS}$ as a function of $h_L=h_1$. There are two special values for $h_L$.  At $h_L=3$, where $\beta_L h_L=\beta_R h_R$, all quantities vanish (equilibrium state). At $h_L=h_R=2$, $\dot Q_L^{\rm NESS}=-\dot Q_R^{\rm NESS}$ and thus $\dot W^{\rm NESS}=0$ (non-driven steady state).}
\label{TwobathNess}
\end{center}
\end{figure}

{\it XX chain coupled to two baths}: Consider a hot left and a cold
right heat baths ($\beta_L<\beta_R$) connected by an XX spin 1/2 chain
with the Hamiltonian in Eq.(\ref{XYH}) with $J_x=J_y=J$. The NESS in
the special case of a uniform magnetic field was analyzed in~\cite{Karevski-Platini}. 
The power and heat from the reservoir to the system are given by Eqs.(\ref{QXY},\ref{WXY}). 
In Figure~\ref{TwobathNess}, we plot $\dot Q_L$, $\dot Q_R$, $\dot W=\dot W_L+\dot W_R$ and
$d_{\rm i}S/dt$ in the NESS as functions of $h_L$. We observe that the heat flows can change sign and that for $h_R=h_L$, they have opposite signs, i.e., $\dot Q_L+\dot Q_R=0$, which means that $\dot W=0$.
We also observe in Figure~\ref{TwobathNess} that $d_{\rm i}S/dt\geq 0$ and vanishes only when $\beta_R h_R=\beta_L h_L$, that is, the second law holds even when heat flows from cold to hot, as is the case
for $h_L>\beta_R h_R/\beta_L$, a situation that would appear to be a
contradiction to the Clausius statement of the second law if we do not realize the presence of $\dot W$.

The previous numerical study of boundary-driven spin chains can be complemented with exact results for power and heat in a two-site boundary-driven spin chain obtained from a full analytical solution
of the NESS (see methods). 
In the NESS, the expression for power Eq.(\ref{WXY}) and heat Eq.(\ref{QXY}) can be written in terms of the spin current~\cite{prosen08}
\[
j_s=4\lambda \,\frac{4J^2(M_R-M_L)}{(h_L-h_R)^2+16J^2+16\lambda^2}
\] 
as $\dot Q_L=-h_Lj_s$, $\dot Q_R=h_Rj_s$ and $\dot W=(h_L-h_R)j_s$.
Thus, for $h_L=h_R$, there is no power, but as the previous expression shows, this does not mean that the spin current vanishes. Moreover, the entropy production rate in the NESS  is
\[
\frac{d_{\rm i}S}{dt}
= (\beta_L h_L-\beta_R h_R) j_s
\]
i.e., the spin current $j_s$ and the affinity $(\beta_L h_L-\beta_R h_R)$ characterize the rate of entropy production in the NESS,
and because $M_r=-\tanh(\beta_r h_r/2)$, the sign of the entropy production rate is given by $(\beta_L h_L-\beta_R h_R)(\tanh(\beta_L h_L/2)-\tanh(\beta_R h_R/2))\geq 0$, where the equality holds only if $\beta_L h_L=\beta_R h_R$. Let us end this analysis by noting that for $\beta_L\leq \beta_R$, this system behaves as 
a heat engine for $\beta_L/\beta_R<h_R/h_L<1$ with efficiency
$\eta\equiv -\dot W/\dot Q_L=1-h_R/h_L \leq
1-\beta_L/\beta_R\equiv\eta_C$, as a refrigerator for
$h_R/h_L<\beta_L/\beta_R<1$ with efficiency $\eta^F\equiv \dot
Q_R/\dot W=1/(h_L/h_R-1)<1/(\beta_L/\beta_R-1)\equiv \eta^F_C$ and as
a heater for $h_R/h_L>1$. Note that the efficiencies are independent
of temperature. These are steady-state operating engines analogous to
those in~\cite{campisi-engine}. 

\section*{Discussion}

\label{secConclu}
In conclusion, the repeated interaction scheme provides a physical description of a system interacting with an environment that, in an appropriate limit,
provides a boundary-driven Lindblad equation for the system. The
Lindblad operators that appear in this equation are determined by the
interaction of the system with the environment, the Hamiltonian of the
copies that form the bath and, importantly, by the fact that it is
refreshed constantly. By computing the thermodynamical quantities for
the full system plus the environment, one can derive the corresponding expressions for the boundary-driven model. One important observation is that due to the refreshing of the reservoir, work is done or extracted 
by the external agent in charge of this refreshing. 
This power drives the system out of equilibrium. Note that this power appears even if the system Hamiltonian and Lindblad operators are time independent. 
We applied our results to spin chains.  
In the single bath case, we found that an XX spin chain with a homogeneous magnetic field relaxes to thermal equilibrium, i.e., a state with zero entropy production, 
while an XY spin chain reaches a driven NESS, a state with a non-zero entropy production $d_{\rm i}S/dt=\beta \dot W>0.$ 
In the two heat bath case, the XX chain for different temperatures
$\beta_R\neq\beta_L$ and a homogeneous magnetic field
reaches a non-driven $\dot W=0$ NESS and an equilibrium state for $\beta_L h_L=\beta_R h_R$ where the entropy production rate,
power, heat flows and spin currents vanish.  
For inhomogeneous magnetic fields, the chain reaches a driven $\dot W\neq 0$ NESS. 
Jumping to a broader context, this work shows that the knowledge of a Lindblad equation for an open system does not determine the heat flows or other thermodynamical
quantities. These quantities also depend on the properties of the environment and how the system is coupled to it. 
Here, we have obtained appropriate expressions for heat flows and power for interactions with an environment of a type recently implemented in a laboratory~\cite{ion-trap}.
But when the reservoir is weakly and passively coupled to the system, i.e. there is no work cost in achieving the coupling, the system is appropriately  described by  
a global~\cite{kosloff} Lindblad equation and the thermodynamical quantities by Eq.(\ref{WQwcl}).
Finally, this work is also an extension of quantum thermodynamics to a class of open quantum systems without local-detailed-balance.

\section*{Methods}

We provide here some details of the calculations mentioned in the main text.

\subsection*{Work, heat and boundary-driven Lindblad equation from the repeated interaction scheme}

For completeness we derive Eq.(\ref{BD2R}) and Eq.(\ref{dotQ}) of the main text. 
Consider $\Delta \rho_S(n\tau)\equiv\rho_S(n\tau+\tau)-\rho_S(n\tau)$. We have from Eq.(\ref{rint}) of the main text
that
\begin{equation}
\Delta \rho_S(n\tau)={\rm Tr}_E[U\rho_S(n\tau)\otimes \rho_E U^\dag-\rho_S(n\tau)\otimes \rho_E ]
\label{iter}
\end{equation}
where we have dropped the label $n$ from $U_n$ and $\rho_n$ in Eq.(\ref{rint})
because the copies are identical and the interaction $V^n=\sum_rV^n_r$ is always of the same form.
The trace ${\rm Tr}_n$ over the state $\rho_n=\rho_E=\omega_{\beta_L}\otimes\omega_{\beta_R}$ is denoted ${\rm Tr}_E$.
The unitary $U=e^{-i\tau(H_S+H_L+H_R+V)}$ in (\ref{iter})
is expanded for small $\tau$ considering the scaling $V=v/\sqrt{\tau}$ and $H_0=H_S+H_L+H_R$
\begin{equation}
U=I-iv\tau^{\frac{1}{2}}-(iH_0+\frac{v^2}{2})\tau-\frac{1}{2}\left\{H_0,v\right\}\tau^{\frac{3}{2}}+{\mathcal O}(\tau^{2}).
\label{Uexp}
\end{equation}
Now, because ${\rm Tr}_E[v\rho_E]=0$ the leading order in the right hand side of (\ref{iter}) is ${\mathcal O}(\tau)$. 
Thus, we divide by $\tau$ and take the limit $\tau\to 0$ and $n\to \infty$ such that $t=n\tau$ and obtain
\[
\dot \rho_S=-i[H_S,\rho_S]+{\rm Tr}_E[v(\rho_S\otimes \rho_E)v]-\frac{1}{2}{\rm Tr}_E\{v^2,\rho_S\otimes \rho_E\}
\]
where the equality ${\rm Tr}_E([H_E,\rho_E])=0$ was used. 

Now we use ${\rm Tr}_r$ to denote the trace over the $r=L$ or $r=R$ system and ${\rm Tr}$ the full trace. 
Because $v=\sum_r v_r$ and ${\rm Tr}_r[v_r\omega_{\beta_r}]=0$, it is possible to split the last two terms in contributions for each reservoir giving Eq.(\ref{BD2R}) in the text:
\begin{equation}
\dot \rho_S=-i[H_S,\rho_S]+\sum_r{\mathcal D}_r(\rho_S)
\end{equation}
with ${\mathcal D}_r(\rho_S)={\rm Tr}_{r}[v_r(\rho_S\otimes \omega_{\beta_r})v_r]-\frac{1}{2}{\rm Tr}_r\{v_r^2,\rho_S\otimes \omega_{\beta_r}\}$. 

We continue with the derivation of Eq.(\ref{dotQ}) of the main text. Let us start from 
$\Delta Q_r={\rm Tr}(H_r^n [\rho_n-\rho_{n}'])$, i.e. Eq.(\ref{QRI}), where $\rho'_{n}={\rm Tr}_S(U_n\rho_{S}([n-1]\tau)\otimes \rho_n U_n^\dag)$. 
Dropping as before the label $n$, in the limit $V\to v/\sqrt{\tau}$ and $\tau\to 0$ we can replace $U$ by (\ref{Uexp}). 
The leading order of $\Delta Q_r$ is ${\mathcal O}(\tau)$
\[
\Delta Q_r=-\tau{\rm Tr}\left((v_rH_rv_r-\frac{1}{2}\{v_r^2,H_r\})\rho_S([n-1]\tau)\otimes\rho_E\right)
\]
or ($\dot Q_r=\Delta Q_r/\tau$)
\begin{equation}
\dot Q_r=-{\rm Tr}\left((v_rH_rv_r-\frac{1}{2}\{v_r^2,H_r\})\rho_S(t)\otimes\omega_{\beta_r}\right)
\label{aQ}
\end{equation}

Consider Eq.(\ref{WRI}) now i.e. $\Delta W_r=-{\rm Tr}(V^n_rU_n\rho_S([n-1]\tau)\otimes \rho_{n}U_n^\dag)$. As before we drop the label $n$.
The leading order is also ${\mathcal O}(\tau)$ but we need $U$ up to ${\mathcal O}(\tau^{3/2})$because $V$ is ${\mathcal O}(1/\sqrt{\tau})$,  ${\rm Tr_E}(V \rho_E)=0$ and ${\rm Tr_E}([(H_S+H_E),V] \rho_S\otimes \rho_E)=0$. 
We obtain
\[
\Delta W_{r}=\tau{\rm Tr}\left((v_r(H_S+H_r)v_r-\frac{1}{2}\{v_r^2,H_S+H_r\})\rho_S([n-1]\tau)\otimes \omega_{\beta_r}\right)
\]
or $\dot W_r=\Delta W_{r}/\tau$
\begin{equation}
\dot W_r={\rm Tr}\left((v_r(H_S+H_r)v_r-\frac{1}{2}\{v_r^2,H_S+H_r\})\rho_S(t)\otimes \omega_{\beta_r}\right)
\label{dotW}
\end{equation}
Expressions (\ref{aQ}) and (\ref{dotW}) correspond to those in Eq.(\ref{dotQ}) from the main text.

\subsection*{The two spin XX chain with inhomogeneous magnetic field}

Consider a XX two sites spin chain and the corresponding Lindblad dynamics Eq.(\ref{Lind}) with $H_S$ given by Eq.(\ref{XYH}) main text (with $J_x=J_y=J$, $h_1=h_L$ and $h_2=h_R$) and the Lindblad dissipator 
\[
{\mathcal D}_r(\rho)=\sum_{\mu\in\{+,-\}}\gamma^{\mu}_r[2L^\mu_r\rho L^{\mu\dag}_r-\{L^{\mu\dag}_r L^\mu_r,\rho\}]
\]
with $\gamma^\pm_r=\lambda_r(1\pm M_r), L^{\pm}_L=\sigma_1^{\pm}$ and $L_R^{\pm}=\sigma_2^{\pm}$ where $\sigma^\pm_j\equiv(\sigma^x_j\pm i\sigma^y_j)/2$.
This system is fully characterized by the correlation functions $\langle \sigma_1^z\rangle_t$, $\langle \sigma_2^z\rangle_t$,
$Y\equiv i\langle \sigma_1^y\sigma_2^x-\sigma_1^x\sigma_2^y\rangle_t$ and $X\equiv \langle \sigma_1^x\sigma_2^x+\sigma_1^y\sigma_2^y\rangle_t$ where $\langle\cdot\rangle_t={\rm Tr}_S(\cdot \rho_S(t))$. They satisfy a close system of equations:
\begin{equation}
\frac{dX}{dt}=-4 \lambda X-i(h_R-h_L)Y 
\label{eqX}
\end{equation}
\begin{equation}
\frac{d\langle\sigma_1^z\rangle_t}{dt}=4 \lambda (M_L-\langle\sigma_1^z\rangle_t)+2iJY 
\label{s1z}
\end{equation}
\begin{equation}
\frac{d\langle\sigma_2^z\rangle_t}{dt}=4 \lambda (M_R-\langle\sigma_2^z\rangle_t)-2iJY 
\label{s2z}
\end{equation}
\begin{equation}
\frac{dY}{dt}=i(h_L-h_R)X-4iJ(\langle\sigma_2^z\rangle_t-\langle\sigma_1^z\rangle_t)-4\lambda Y
\label{eqY}
\end{equation}

From Eqs.(\ref{QXY} and \ref{WXY}) in the main text we note that $\dot W_L=2\lambda J X=\dot W_R$, while the first term in the right hand side of (\ref{s1z}) is $2\dot Q_L/h_L$ and the corresponding term in (\ref{s2z}) is $2\dot Q_R/h_R$.
Moreover the spin current~\cite{prosen08} is $j_s=iJY$. In the steady state the left-hand-side of the system (\ref{eqX},\ref{s1z},\ref{s2z},\ref{eqY}) vanishes and $\dot W=\dot W_L+\dot W_R=(h_L-h_R)j_s$, $\dot Q_L=-h_Lj_s$ and $\dot Q_R=h_Rj_s$. The current given in the main text
is obtained by solving the full system in the NESS.

\subsection*{Heat from a given reservoir in the weak-coupling limit}

Consider a system coupled to several reservoirs as discussed in {\it ``universe" under unitary evolution}. The heat that comes from one of them, for instance the $r=L$ reservoir is $Q_L\equiv {\rm Tr}[H_L(\rho_{\rm tot}(0)-\rho_{\rm tot}(t))]$.
The methods developed in full counting statistics~\cite{esposito-RMP} gives
$Q_L=i\partial_\lambda G(\lambda)|_{\lambda=0}=-i{\rm Tr}(t\partial_\lambda {\mathcal L}_{\lambda}\rho_S(t))|_{\lambda=0}$ 
where ${\mathcal L}_{\lambda}\rho=-i(H_{\rm tot}^\lambda\rho-\rho H_{\rm tot}^{-\lambda})$ is a modified evolution super-operator with $H_{\rm tot}\to H_{\rm tot}^\lambda=e^{-i(\lambda/2) H_L}H_{\rm tot}e^{i(\lambda/2) H_L}$.
When this modification is done for a system in the weak coupling Born-Markov-secular approximation one obtain~\cite{sawaga-hayakawa,finlandia}  ${\mathcal L}_{\lambda} \rho=-i[H_S,\rho]+\sum_r{\mathcal D}_{r,\lambda} \rho$ where
only the dissipator associated to the $r=L$ reservoir depends on $\lambda$ as
\begin{equation}
{\mathcal D}_{L,\lambda} Y=\sum_{l}\sum_{\omega}h^+_{l}(\omega)
(e^{i\lambda \omega}2A_{l}^{\dag\omega}YA^{\omega}_{l}-\{A_l^\omega A^{\dag\omega}_{l},Y\})
+h^-_{l}(\omega)e^{-i\lambda\omega}(2A_{l}^{\omega}YA^{\dag\omega}_{l}-\{A_l^{\dag\omega}A^{\omega}_{l},Y\}).
\label{WCL}
\end{equation}
Here $A_l^\omega$ are system eigen-operators obtained from the coupling of the system to the left reservoir~\cite{BreuerBook,sawaga-hayakawa,finlandia} and  $h^+_{l}(\omega)=e^{-\beta_L\omega}h^-_{l}(\omega)$. 
A slow time dependence of the system can be included, see~\cite{sawaga-hayakawa}.
From Eq.(\ref{WCL}) we obtain
\begin{equation}
-i(\partial_\lambda {\mathcal D}_{L,\lambda})|_{\lambda=0}Y=2\sum_{l,\omega}\omega \left(h^+_{l}(\omega)
A_{l}^{\dag\omega}YA^{\omega}_{l}
-h^-_{l}(\omega)A_{l}^{\omega}YA^{\dag\omega}_{l}\right).
\label{aux1}
\end{equation}
Thus $\dot Q_L=-i{\rm Tr}[(\partial_\lambda {\mathcal D}_{L,\lambda})\rho_S(t)]|_{\lambda=0}$ where we used that in this limit the dynamics is Markovian.
We have to compare this with the heat flow defined in section {\it ``open system weakly and passively coupled to the environment"}, $\dot Q_L(t)={\rm tr}(H_S{\mathcal D}_L(\rho_S))={\rm tr}(\rho_S{\mathcal D}^\dag_L(H_S))$, where the dissipator ${\mathcal D}_L$
in the same weak coupling Born-Markov-secular approximation is given by ${\mathcal D}_{L,\lambda=0}$, from which we compute
\begin{equation}
{\mathcal D}_L^\dag(H_S)=2\sum_{l,\omega}\omega \left(h_{l}^+(\omega)A_{l}^\omega A_l^{\dag\omega}-h_{l}^-(\omega)A_l^{\dag\omega}A_{l}^\omega\right).
\label{aux2}
\end{equation}
To obtain this we used $[H_S,A^{\omega\dag}_lA^{\omega}_{l'}]=0$~\cite{BreuerBook}. Taking the trace in Eq.(\ref{aux1}) and in Eq.(\ref{aux2}) the desired equality $-i{\rm Tr}[(\partial_\lambda {\mathcal D}_{L,\lambda})\rho_S(t)]|_{\lambda=0}={\rm tr}(H_S{\mathcal D}_L(\rho_S))$ is found. 
Now, since the heat flow to a system weakly and passively coupled to the $L$ heat-bath is given by  $\dot Q_L(t)={\rm tr}(H_S{\mathcal D}_L(\rho_S))$, the corresponding definition for work follows and the entropy production given in Eq.(\ref{eprod0}) as well.



\section*{Acknowledgements}

This research is funded by  Fondecyt grant 1151390.






\end{document}